\documentclass{article}
\pdfoutput=1
\usepackage{amssymb}
\usepackage{mathtools}

\usepackage{bm}
\usepackage{bbm}
\usepackage[T1]{fontenc}
\usepackage[utf8]{inputenc}

\usepackage[table]{xcolor}
\usepackage{tabu}
\usepackage{array}

\usepackage{natbib}

\usepackage{tikz}
\usepackage{units}
\usepackage{ifthen}
\usetikzlibrary{calc}

\title{Learning meets Assessment: On the relation between Item Response Theory and Bayesian Knowledge Tracing}
\author{Benjamin Deonovic, Michael Yudelson, Maria Bolsinova, \\
Meirav Attali, and Gunter Maris}

\begin{document}
\maketitle
\begin{abstract}
Few models have been more ubiquitous in their respective fields than Bayesian knowledge tracing and item response theory. Both of these models were developed to analyze data on learners. However, the study designs that these models are designed for differ; Bayesian knowledge tracing is designed to analyze longitudinal data while item response theory is built for cross-sectional data. This paper illustrates a fundamental connection between these two models. Specifically, the stationary distribution of the latent variable and the observed response variable in Bayesian knowledge Tracing are related to an item response theory model. This connection between these two models highlights a key missing component: the role of education in these models. A research agenda is outlined which answers how to move forward with modeling learner data. 
\end{abstract}

\section{Introduction}
Learning and assessment deal with related, yet distinct concepts. Learning can be defined as the acquisition of knowledge, skills, values, beliefs, and habits through experience, study, or instruction. Assessments are instruments designed to observe behavior in a learner and produce data that can be used to draw inference about the knowledge, skills, values, beliefs, and habits that the learner has. Although learning and assessment are both key to education, the statistical models used to describe learning data and assessment data have significantly diverged and grown to leverage the salient features and distinct assumptions that are embodied in their respective data sets. The fields of educational data mining and learning analytics harness the dynamic, temporal, and large scale nature of learning data to construct models which can be used to predict learner performance, personalize and adapt instructional content, recommend intervention and curriculum changes, and provide information visualization to track progress. On the other hand, the same objectives are targeted by the field of psychometrics, using cross-sectional assessment data rather than longitudinal data. Specifically this paper will explore the connections between Bayesian knowledge Tracing (BKT) and item response theory (IRT). BKT, a statistical model in educational data mining, is the most ubiquitous model used for data obtained from intelligent tutoring systems, which are systems constructed to provide immediate and customized instruction to learners. IRT, a modeling framework developed in the field of psychometrics, was designed for constructing and analyzing assessments.

Historically, the research in BKT and IRT models has had little overlap, as on the surface these models seem to be completely different and incompatible. Both of these models fit their respective data sets well, but each has flaws. Due to the relationship between the longitudinal learning data and the cross-sectional assessment data, we posit that there exists a relationship between BKT and IRT models. Indeed we will show that there is an intimate connection between these two models that places BKT and IRT under an umbrella of general models of learning and assessment data. First in Section \ref{sec:BKT} the BKT model is explained in detail and extensions to the standard model that have been described in the literature are also listed. Section \ref{sec:IRT} describes the IRT model and its extensions. Section \ref{sec:crit} discusses the shortcomings of the respective models in the context of learning. Section \ref{sec:bridge} describes the connection between the BKT models and IRT models and how the shortcomings of each model can be addressed by incorporating concepts from the other. 

It should be noted that this paper does not describe a novel statistical model nor an algorithm to fit a statistical model to either the longitudinal learning data or cross-sectional assessment data. Rather this paper identifies a key theoretical connection between two existing and popular models. However, this result is not inconsequential. The connection between BKT and IRT highlights that there is a crucial ingredient missing from both. That crucial ingredient is education. Only when learning, assessment, and education go hand in hand can there be hope to make progress. Hence in Section \ref{sec:future} we end this paper with sketching a research agenda for achieving this.

\section{BKT} \label{sec:BKT}
Bayesian Knowledge Tracing or BKT \citep{corbett1994} is a modeling paradigm frequently used in the field of Intelligent Tutoring Systems (ITS) where it is tasked with continuously tracking the process of student knowledge acquisition and serves as the basis for selecting the next problem set or skill that a student should work on, once mastery has been attained on the current problem set or skill.  In BKT, skills are modeled as (latent) binary variables (mastered/not-mastered) and learning is characterized as a transition between these states. Let $Z_{pkt}$ denote the dichotomous state (i.e. mastery/not-mastery) of the $k$th skill for the $p$th person at attempt $t$ for $k=1,\ldots,K$, $p=1,\ldots,n$ and $t=1,\ldots,T_p$. Originally, BKT was developed with cognitive theory of learning in mind. Each model addresses one skill and assumes that it is relatively fine-grained (e.g., addition, subtraction, division) \citep{corbett1994}. Granularity of the skills is a subject of experimental research and, for example, ‘addition’ could be split to ’single-digit addition’ and ’multi-digit addition’ if  the data indicates the split is warranted \citep{koedinger2013}.

The standard BKT model utilizes five global (i.e.\ shared among all individuals in the model) parameters per skill $\bm\pi_k = (\pi_{0k},\pi_{{\ell}k},\pi_{{\phi}k},\pi_{sk}, \pi_{gk})^\intercal$. The probability of learning skill $k$, after completing an item utilizing that skill, is denoted by $\pi_{{\ell}k}$, the probability of forgetting skill $k$ once learned is denoted by $\pi_{{\phi}k}$, the probability of an incorrect answer on an item when skill $k$ is mastered (a slip) is denoted by $\pi_{sk}$, the probability of a correct answer when skill $k$ is unmastered (a guess) is denoted by $\pi_{gk}$, and $\pi_{0k}$ is the probability that a learner is in the mastery state for that skill before beginning the assessment at hand. Note for the standard model these parameters are not individual specific, each skill is assumed to be independent of the other skills, and once a skill is learned it cannot be forgotten (i.e. $\pi_{{\phi}k}=0$). The BKT model is equivalent to a two state hidden Markov model (HMM) (see Figure \ref{fig:hmm}), where $Z_{pkt}$ represents the hidden or latent state, with a dichotomous emission $X_{pkt}$. The notation and description used in the original \citet{corbett1994} paper is listed in Table \ref{tab:t1}. 

\begin{table}[t]
    \centering
    \tabulinesep=1.2mm
    \rowcolors{1}{gray!30}{white}
    \begin{tabu} to \textwidth {cclX[3,l,m]}
        $\pi_{0k}$& $p(L_0)$ & Initial Learning & the probability a skill is in the learned state prior to the first opportunity to apply the skill \\
        $\pi_{{\ell}k}$ & $p(T)$ & Acquisition & the probability a skill will make the transition from the unlearned to the learned state following an opportunity to apply the skill\\
        $\pi_{gk}$ & $p(G)$ & Guess & the probability a student will guess correctly if a skill is in the unlearned state\\
        $\pi_{sk}$ & $p(S)$ & Slip & the probability a student will slip (make a mistake) if a skill is in the learned state\\
        $\pi_{{\phi}k}$ &  & Forgetting & the probability a skill will make the transition from the learned to the unlearned state following an opportunity to apply the skill
    \end{tabu}
    \caption{The learning and performance parameters as described in the original \citet{corbett1994} paper along with the original notation.}
    \label{tab:t1}
\end{table}

\tikzstyle{state}=[shape=circle,draw=blue!50,fill=blue!20]
\tikzstyle{observation}=[shape=rectangle,draw=orange!50,fill=orange!20]
\tikzstyle{lightedge}=[<-,dotted]
\tikzstyle{mainstate}=[state,thick]
\tikzstyle{mainedge}=[<-,thick]
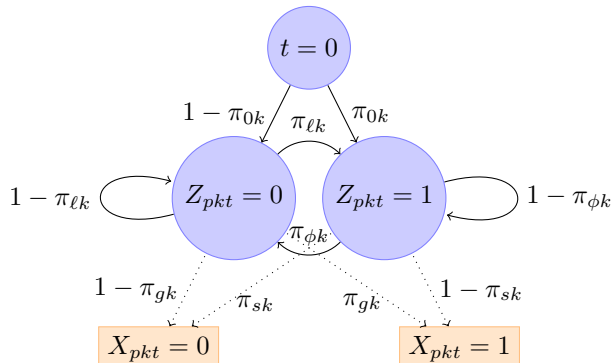
\begin{figure}
\centering
\begin{tikzpicture}[]
\node[state] (s1) at (0,2) {$Z_{pkt}=0$}
    (s1) edge [loop left] node {$1-\pi_{{\ell}k}$}  (s1);
\node[state] (s2) at (2,2) {$Z_{pkt}=1$}
    edge [<-,bend right=45] node[auto,swap] {$\pi_{{\ell}k}$} (s1)
    edge [->,bend left=45]  node[auto,swap] {$\pi_{{\phi}k}$} (s1)
    (s2) edge [loop right] node {$1-\pi_{{\phi}k}$} (s2);
\node[state] (start) at (1,4) {$t=0$}
    edge [->] node[left] {$1-\pi_{0k}$} (s1)
    edge [->] node[right] {$\pi_{0k}$} (s2);
\node[observation] (y1) at (-1,0) {$X_{pkt}=0$}
    edge [lightedge] node[left] {$1-\pi_{gk}$} (s1)
    edge [lightedge] node[right, pos=0.25] {$\pi_{sk}$} (s2);
\node[observation] (y2) at (3,0) {$X_{pkt}=1$}
    edge [lightedge] node[left, pos=0.25] {$\pi_{gk}$} (s1)
    edge [lightedge] node[right] {$1-\pi_{sk}$} (s2);
\end{tikzpicture}
\caption{The transition and emission probabilities of the standard BKT model with forgetting. It is equivalent to a two state hidden Markov model, where $Z_{pkt}$ represents the hidden or latent state for the $k$th skill of person $p$ at time $t$, with a dichotomous emission $X_{pkt}$ representing the observed response (correct/incorrect) for the $p$th person of the $k$th skill at attempt $t$.}
\label{fig:hmm}
\end{figure}

The standard BKT model only utilizes longitudinal performance data and does not permit features such as student specific or item specific parameters. A plethora of extensions to BKT have been developed, including variants that measure the effect of students' individual characteristics \citep{pardos2011, lee2012,  yudelson2013, khajah2014A, khajah2014B}, assessing the effect of help in a tutor system \citep{beck2008, sao2013}, controlling for item difficulty \citep{gowda2010, pardos2011, schultz2013}, measuring impact of time between attempts \citep{qiu2010}, incorporating forgetting \citep{nedungadi2015, khajah2016}, and measuring the effect of subskills \citep{xu2010}. 

\section{IRT} \label{sec:IRT}

Statistical models utilizing IRT have played an extensive role in assessment and educational measurement. The history of IRT can be traced back to pioneering work by Louis Thurstone in the 1920s, and seminal work by Bert Green, Alan Birnbaum, Frederic Lord, and Georg Rasch in the 1950s and 60s \citep{green50A,lord1951,birnbaum1967,rasch1960}. The IRT model consists of three basic assumptions: 1) the probability that a person correctly answers an item follows a specific parameteric functional form called the item characteristic curve (ICC) or item response function (IRF), which depends on parameter(s) for that person and parameter(s) for the item; 2) this IRF is monotonically increasing function with respect to a person's ability; and 3) given the person's ability, the items are considered conditionally independent. One specific IRT model is the four parameter logistic (4PL) model which models the probability that individual $p$ answers item $i$ correctly by the logistic function 

\begin{equation}
P(Y_{pi}=1 | \theta_p, a_i, b_i, c_i, d_i)=c_i+(d_i - c_i)\dfrac{\exp{\left[a_i(\theta_p - b_i)\right]}}{1+\exp{\left[a_i(\theta_p - b_i)\right]}}.
\end{equation}

Where $Y_{pi}$ is the response of person $p$ on item $i$, $\theta_p$ represents the person's ability, $b_i$ represents an item's difficulty, $a_i$ represents an item's discrimination, $c_i$ represents guessing, and $d_i$ represents inattention (slips). Subsets of the 4PL include the 3PL ($d_i=1$), 2PL ($d_i=1, c_i=0$), and 1PL ($d_i=1, c_i=0, a_i=1$) which is often refereed to as the Rasch model.

\begin{figure}
\includegraphics[width=\textwidth]{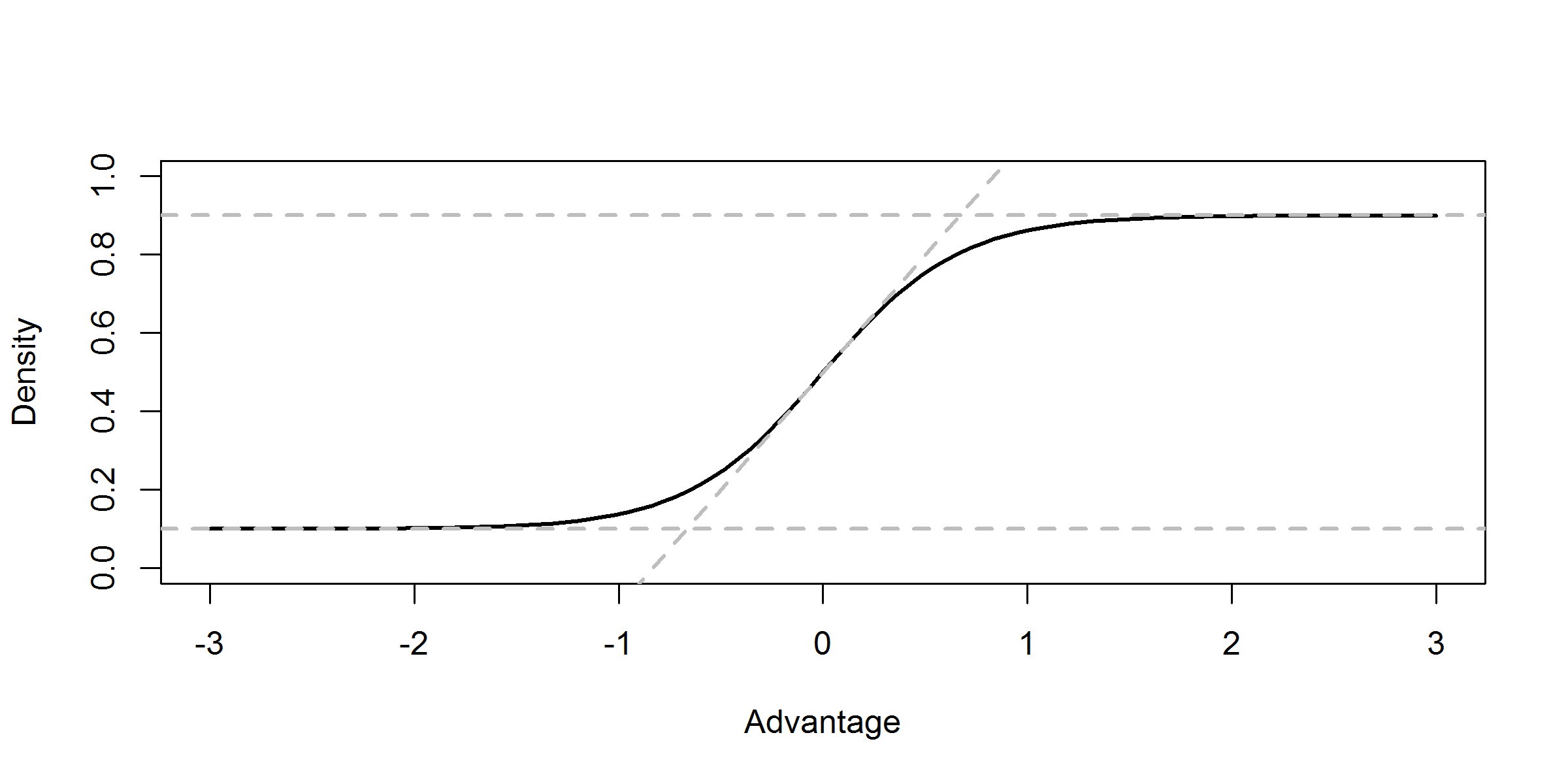}
\caption{Item response function (IRF) of the 4PL model. The horizontal axis is the advantage, $\theta_p - b_i$. The IRF is the black solid line, the upper gray dashed line is the upper asymptote $d_i$, the lower gray dashed line is the lower asymptote $c_i$, and the last gray dashed line is the maximum slope of the IRF which is $a_i(d_i-c_i)/4$.}%
\end{figure}

Originally, the first IRT models that were developed treated ability as a static, unidimensional parameter. This framework made IRT especially well-suited for ranking a set of individuals by their ability, hence its use in assessment, particularly summative assessments. However, these assumptions make IRT inappropriate for analyzing data that are not cross-sectional in nature, such as data collected by continuous assessments, or sometimes formative assessments. Historically the field of psychometrics has been mostly concerned with the analysis of cross-sectional data, such as summative assessments, while the fields of educational data mining and learning analytics have focused on longitudinal data such as data produced by continuous and formative assessments.

The research on IRT is extensive and thorough. Countless extensions and generalizations of the 4PL and other IRT models have been developed, too many to list and cite here. For a review of IRT and its extensions see \citet{irthandbook}. Notably, for the context of this paper, three extensions of IRT which make it more amenable to learning data are briefly described: adaptive item administration, multidimensionality of ability, and time-varying ability. 

First, traditional assessments consist of fixed forms comprised of a set of items. This may not be appropriate in a learning context in which the strengths and weakness of different learners can be harnessed to present the learner with more appropriate items for their skill level. One way to administer an adaptive set of items, rather than a fixed form, proposed by \citet{lord1977}, is to administer an item which would maximize the Fisher information given the current estimate of a learner's ability. Selecting the item which maximizes the Fisher information is an efficient way to select an item as maximizing the Fisher information is equivalent to minimizing the lower bound of the variance of the ability estimate. This method serves the goal of testing, which is to efficiently estimate an individual’s ability, however requires a large pool of items.

Although adaptive item administration is a key feature of a learning system, models which utilize a unidimensional representation of ability, such as the standard IRT model, are incapable of inferring what aspects of the material the individual has mastered or not mastered. Multidimensional IRT (MIRT) extends IRT to allow for a multidimensional ability parameter \citep{reckase1972}. In the multidimensional 4PL (MD-4PL) the ability parameter and the discrimination parameter are expanded to be $k$ dimensional $\bm{\theta}_p = (\theta_{p1},\ldots,\theta_{pk})^\intercal$ and $\mathbf{a}_i = (a_{i1},\ldots,a_{ik})^\intercal$ with the following IRF 

\begin{equation}\label{eq2}
P(Y_{pi}=1 | \bm{\theta}_p, \mathbf{a}_i, \beta_i, c_i, d_i)=c_i+(d_i - c_i)\dfrac{\exp{(\mathbf{a}_i^\intercal\bm\theta_p + \beta_i)}}{1+\exp{(\mathbf{a}_i^\intercal\bm\theta_p + \beta_i)}}. 
\end{equation}

In Equation \ref{eq2} $\beta_i$ is labeled an item intercept and is related to the item difficulty. This model is able to estimate a more complex construct rather than just a unidimensional estimate of a learners ability. It is important to note that this formulation of MIRT is compensatory. This means that a high ability in one dimension of the multidimensional $\bm{\theta}_p$ can compensate for low ability in the other dimensions. Non-compensatory MIRT have also been defined \citep{sympson1978}. 

Finally, several extensions to traditional IRT have been proposed to allow for ability to be time dependent (to be able to fit longitudinal learning data). Extensions that have been published include extensions of MIRT that allow for longitudinal data \citep{embretson1991}; state space modeling approaches that have been used to model attitudinal changes \citep{martin2002} and growth in reading ability \citep{WangXJ2013}; a deterministic moment-matching method to estimate dynamic ability with real-time continuously streaming data \citep{weng2017}; and a multidimensional state-space approach \citep{ekanadham2017}. Additionally, two comprehensive theses have been written which introduce several dynamic IRT models \citep{rijn2008,studer2012}. A simple example of a dynamic IRT model (in the state-space modeling approach) is as follows: 

\begin{align}
\text{System equation: }&\quad\theta_{p,t} = \theta_{p,t-1} + \epsilon_{p,t}\\
\text{Observation equation: }&\quad P(Y_{pit}=1|\theta_{pt},b_i) = \dfrac{\exp(\theta_{pt} - b_i)}{1 + \exp(\theta_{pt}-b_i)}.
\end{align} 

This is a dynamic extension of the 1PL model, where $\theta_{pt}$ represents person $p$'s ability at time $t$. The various models mentioned in the literature above vary on the specific form of these equations and how the parameters are estimated. 

Another extension of IRT that accounts for longitudinal data comes from the learning data modeling literature. These extensions include the Additive Factor Analysis (AFM) \citep{cen2006} and the Performance Factor Analysis (PFA) \citep{pavlik2009} models. These models extend IRT to longitudional data by dropping the requirement of conditional independence for the same items. Instead the dependence is modeled by linear factors involving the number of attempts on the item along with other factors. These models and more complex versions have been shown \citep{maclellan2015} to consistently outperform BKT across 5 datasets in better prediction via cross-validation.

\section{Criticisms of BKT and IRT for Learning Systems}
\label{sec:crit}
The core issue with both BKT and IRT is their lack of a placeholder for education in the model. Although the BKT model can estimate the rate at which learning occurs through the parameter $\pi_{{\ell}k}$ and the IRT model is capable of estimating the learning that has occurred (i.e.\  the student's faculties) through the ability parameter $\theta_p$, there is no component in either model to denote teaching or education that is occurring to the learners, nor how differences in teaching lead to differences in learning outcomes (IRT), or the learning process (BKT).

The BKT model is basically a ballistic model, where the learning process is closer to firing a cannon, with the path being almost entirely determined by the initial conditions (i.e., parameters), than it is to flying a plane, with a pilot (i.e.\ education) steering and changing the course of the plane as needed. This is one way in which education interacts with assessment and learning. Education can be seen as setuping up a canon (e.g.\ a system powered by BKT), firing it (i.e.\ having learners go through the system answering questions and fitting the BKT parameters), seeing where the cannon ball lands (i.e.\ interpreting and analyzing the resulting BKT parameters), and then reconfiguring the system and doing it all over to optimize some criterion. We believe this process should be made more holistic, with the effects of education incorporated into the model. 

IRT models on the other hand are inherently cross-sectional, and aim to explain observed differences in what has been learned. Such models however have little to offer in explaining how these observed differences came into existence, or what measures could reduce or alter them.

Furthermore, there is a requirement that the skills in the BKT model (and assignment/tagging of particular items to skills) needs to be done \textit{apriori}. One could argue that this process serves as a sort of placeholder for education to some extent. However, this aspect is quite removed from the model itself and may be arbitrary. It is possible to forego this issue by considering each unique set of problems to be its own "skill". This can then be used as a proxy mapping for knowledge components and the BKT model can be fit to the data. This allows one to identify the BKT learning rate parameters without skill tags at the expense of external validity (you can't describe what the skills are without consulting content experts) and some degree of overfitting (it's unlikely that you have as many unique skills as unique problem sets in most learning environments).

A similar issue arises in MIRT analysis with the $\mathbf{a}_i$ parameters: should it be specified which $\mathbf{a}_i$ are nonzero beforehand (confirmatory MIRT) or should they all be freely estimated with some identifiability constraint (exploratory MIRT)? These skills are also considered to be independent in BKT, which may not be appropriate, and similarly in MIRT the components are often estimated to be orthogonal to each other but then rotated to obtain some kind of interpretable result.

\section{Connecting BKT and IRT} \label{sec:bridge}
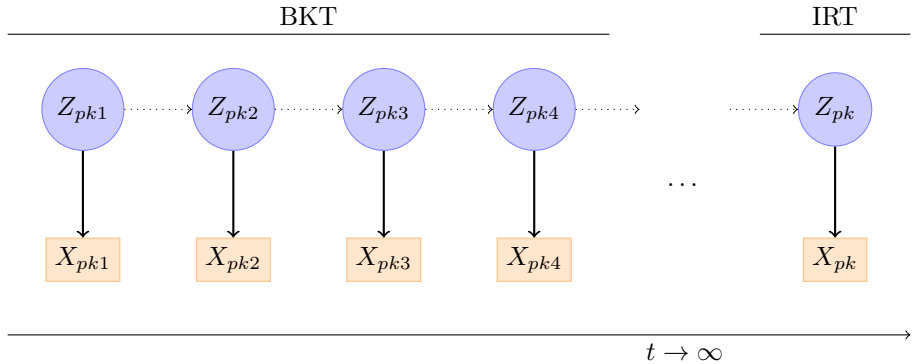
\begin{figure}[t]
\centering
\begin{tikzpicture}[]
  \draw (-1,3) -- node[auto=false, above]{BKT} (7,3);
  \draw (9,3) -- node[auto=false, above]{IRT} (11,3);
  \node[state] (Z1) at (0,2) {$Z_{pk1}$};
  \node[state] (Z2) at (2,2) {$Z_{pk2}$};
  \node[state] (Z3) at (4,2) {$Z_{pk3}$};
  \node[state] (Z4) at (6,2) {$Z_{pk4}$};
  \node[state] (Zinf) at (10,2) {$Z_{pk}$};  
  \node[observation] (X1) at (0,0) {$X_{pk1}$};
  \node[observation] (X2) at (2,0) {$X_{pk2}$};
  \node[observation] (X3) at (4,0) {$X_{pk3}$};
  \node[observation] (X4) at (6,0) {$X_{pk4}$};
  \node[observation] (Xinf) at (10,0) {$X_{pk}$};
  \path[lightedge,->] (Z1) edge node[above] {} (Z2);
  \path[lightedge,->] (Z2) edge node[above] {} (Z3);
  \path[lightedge,->] (Z3) edge node[above] {} (Z4);
  \draw[shorten >=0.6cm, lightedge,->] (Z4) to node[above] {} (8,2);
  \path (6,0) -- node[auto=false]{\ldots} (10,2);
  \draw[shorten <=0.6cm, lightedge,->] (8,2) to node[above] {} (Zinf);
  \path[thick,->] (Z1) edge node[above] {} (X1);
  \path[thick,->] (Z2) edge node[above] {} (X2);
  \path[thick,->] (Z3) edge node[above] {} (X3);
  \path[thick,->] (Z4) edge node[above] {} (X4);
  \path[thick,->] (Zinf) edge node[above] {} (Xinf);+
  \draw [->] (-1,-1) to node[pos=0.75, below]{$t\to\infty$} (11,-1);
\end{tikzpicture}
\caption{The left hand side illustrates the HMM for the BKT data for a specific skill mastery of a single individual. The right hand side illustrates the equilibrium distribution of the latent mastery variable which can be shown to follow an IRT distribution. The fact that IRT is placed at the end of the BKT's state-transition chain serves to indicate that in the limit the distribution of the latent skill in BKT converges to IRT. Also, traditionally, summative tests are often given after learning has occurred. However, note that there is no requirement that precludes administering such a test before or during the targeted period of learning.}
\label{fig:singlehmmtoirt}
\end{figure}

In this section we develop a unified framework encompassing both the standard BKT model and the IRT family of models. To motivate the construction of the unified framework we demonstrate for the BKT model that at equilibrium the distribution of the latent mastery variable and the distribution for a correct response both follow an IRT model. 

Let the transition and emission matrices of the BKT model be denoted by $\mathbf{A}$ and $\mathbf{B}$ respectively

\begin{equation}
\mathbf{A} = \begin{pmatrix}
1 - \pi_{{\phi}k} & \pi_{{\phi}k}\\
\pi_{{\ell}k} & 1 - \pi_{{\ell}k}
\end{pmatrix} \qquad \mathbf{B} = \begin{pmatrix}
1 - \pi_{gk} & \pi_{gk}\\
\pi_{sk} & 1 - \pi_{sk}
\end{pmatrix}.
\end{equation}

Let $\left\{Z_{pkt}\right\}_{t=1,\ldots,T_p}$ denote the Markov chain formed by the transitions of the latent state variable in the BKT model. This Markov chain has a finite state space and is irreducible meaning it is possible to get to any state from any state (as long as $\pi_{{\phi}k}$ and $\pi_{{\ell}k}$ are not equal to zero). Furthermore this Markov chain is aperiodic, meaning every state can be returned to at every time point. Therefore the stationary distribution for this Markov chain, $\bm\lambda^\intercal = (\lambda_0, \lambda_1)$, exists and is the unique solution which satisfies $\bm\lambda = \mathbf{A}^\intercal\bm\lambda$ \citep{ross2014} where $\lambda_0 = P(Z_{pk} = 0)$, $\lambda_1 = P(Z_{pk} = 1)$, and $Z_{pk}$ is the latent mastery variable at equilibrium, i.e.\ a binary random variable for which $Z_{pk}=1$ indicates person $p$ has mastered skill $k$. After some simple algebra this stationary distribution can be shown to be 

\begin{equation}
\bm\lambda^\intercal = \begin{pmatrix} 
\dfrac{\pi_{{\phi}k}}{\pi_{{\ell}k} + \pi_{{\phi}k}}, & 
\dfrac{\pi_{{\ell}k}}{\pi_{{\ell}k} + \pi_{{\phi}k}}
\end{pmatrix}.
\end{equation}

This is similar to a Rasch model, which after reparameterizing $\theta_k = \log\pi_{{\ell}k}$ and $b_k = \log{\pi_{{\phi}k}}$ we get 

\begin{align}
\dfrac{\pi_{{\ell}k}}{\pi_{{\ell}k} + \pi_{{\phi}k}} &= \dfrac{\exp{(\theta_k - b_k)}}{1 + \exp{(\theta_k - b_k)}} \label{eq7}\\
\dfrac{\pi_{{\phi}k}}{\pi_{{\ell}k} + \pi_{{\phi}k}} &= 1 - \dfrac{\exp{(\theta_k - b_k)}}{1 + \exp{(\theta_k - b_k)}}. \label{eq8}
\end{align} 

This is something similar to the 1PL IRT model for the hidden mastery variable. However, note that instead of a person specific ability parameter and an item specific difficulty there is a skill specific ability and skill specific difficulty. It must be noted that the identifiability of BKT parameters has been discussed at length \citep{beck2007,vandesande2013,gweon2015,doroudi2017} and without constraints the parameters are not identifiable, especially if the forgetting parameter is included. This becomes quite evident in Equations \ref{eq7} and \ref{eq8} in which the parameters $\theta_k$ and $b_k$, which are indexed by the same skill, are not identifiable. For now let us put aside this issue and see this connection through to the end. 

Let $X_{pk}$ be the random variable corresponding to a response of person $p$ to an item utilizing skill $k$ when their learning state has reached equilibrium. By adding in the emissions probabilities we obtain

\begin{align}
\begin{split}
P(X_{pk} = 1) &= P(X_{pk} = 1 | Z_{pk}=1)P(Z_{pk}=1) + \\
  &\hspace{1in}P(X_{pk} = 1 | Z_{pk}=0)P(Z_{pk}=0)
\end{split}\\
  &= (1 - \pi_{sk})\dfrac{\pi_{{\ell}k}}{\pi_{{\ell}k} + \pi_{{\phi}k}} + \pi_{gk}\dfrac{\pi_{{\phi}k}}{\pi_{{\ell}k} + \pi_{{\phi}k}}\\
  &= (1 - \pi_{sk})\dfrac{\exp{(\theta_k - b_k)}}{1 + \exp{(\theta_k - b_k)}} + \pi_{gk}\left(1 - \dfrac{\exp{(\theta_k - b_k)}}{1 + \exp{(\theta_k - b_k)}}\right)\\
  &= \pi_{gk} + ((1-\pi_{sk}) - \pi_{gk})\dfrac{\exp{(\theta_k - b_k)}}{1 + \exp{(\theta_k - b_k)}}.
\end{align}

Thus, at equilibrium the BKT model corresponds to a (4 - 1)PL (four minus one parameter logistic) skill-centric IRT model, i.e.\ a 4PL model with discrimination parameter set to 1; item specific difficulty, guessing, and slipping replaced with skill specific quantities; and individual ability replaced with skill specific ability. Note that if we further restricted the BKT guess and slip parameters to be 0 and 1 respectively the resulting equilibrium distribution would be the 1PL IRT model. Any restrictions imposed on the BKT model parameters, for purposes of identifiability for example, that leave the Markov chain formed by the latent variable distributions irreducible and aperiodic will have a corresponding IRT-like model for the equilibrium distribution of the response with analogous restrictions. 

So far we have shown how a limiting distribution associated with BKT is related to a type of IRT model. This model however is quite strange. The parameters in the model are not all identifiable and the model does not separate between person parameters and item parameters. The strength of IRT models lies in separating the effects of individuals and specific items. What sort of hidden Markov structure would then lead to the 4PL IRT model described above in Section \ref{sec:IRT}? We must construct separate HMMs for every learner-item pair. These HMMs are the same as the ones described for BKT, except that the learner-skill latent mastery variable $Z_{pkt}$ is replaced by a learner-item latent mastery variable $W_{pit}$ and at equilibrium the latent mastery variable $Z_{pk}$ is replaced by $W_{pi}$. The emission parameter $X_{pkt}$ is replaced with $Y_{pit}$ and at equilibrium $X_{pk}$ is replaced by $Y_{pi}$. Additionally, the standard BKT parameters $\bm\pi_k = (\pi_{0k},\pi_{{\ell}k},\pi_{{\phi}k},\pi_{sk}, \pi_{gk})^\intercal$ must be replaced by $(\pi_{0i},\pi_{{\ell}p},\pi_{{\phi}i},\pi_{si}, \pi_{gi})^\intercal$, where $\pi_{0k}$ is replaced by an item specific initial probability $\pi_{0i}$, $\pi_{{\ell}k}$ is replaced by a learner specific value $\pi_{{\ell}p}$, $\pi_{{\phi}k}$ is replaced by an item specific forgetting rate $\pi_{{\phi}i}$, and guessing and slipping values are item rather than skill specific. Thus at equilibrium of this new HMM we get the following

\begin{align}
\begin{split}
P(Y_{pi} = 1) &= P(Y_{pi} = 1 | W_{pi}=1)P(W_{pi}=1) + \\
  &\hspace{1in}P(Y_{pi} = 1 | W_{pi}=0)P(W_{pi}=0)
\end{split}\\
  &= (1 - \pi_{si})\dfrac{\pi_{{\ell}p}}{\pi_{{\ell}p} + \pi_{{\phi}i}} + \pi_{gi}\dfrac{\pi_{{\phi}i}}{\pi_{{\ell}p} + \pi_{{\phi}i}}\\
  &= (1 - \pi_{si})\dfrac{\exp{(\theta_p - b_i)}}{1 + \exp{(\theta_p - b_i)}} + \pi_{gi}\left(1 - \dfrac{\exp{(\theta_p - b_i)}}{1 + \exp{(\theta_p - b_i)}}\right)\\
  &= \pi_{gi} + ((1-\pi_{si}) - \pi_{gi})\dfrac{\exp{(\theta_p - b_i)}}{1 + \exp{(\theta_p - b_i)}}. \label{eq16}
\end{align}

Equation \ref{eq16} can be recognized to be almost the 4PL IRT model described in Section \ref{sec:IRT} with discrimination paramter set to 1. Note that since this derivation deals with the equilibrium distribution the initial probability parameter $\pi_{0i}$ was not involved and thus the actual specification of this parameter (and whether it is skill specific or item specific) is inconsequential. 

It should be noted that the above derivations correspond to a "fixed-effects" version of the IRT model. However, the derivations are still valid if instead of a standard HMM we consider a mixed HMM \citep{altman2007}, where the transition parameters are allowed to be drawn from some distribution. This is turn will result in a "random-effects" IRT model. Furthermore, it should be noted that his stationary distribution is different from that obtained from standard BKT model in which the forgetting parameter is constrained to be zero. Because of this constraint the traditional BKT model will have a stationary distribution that is $P(X_{pk}=1) = 1 - \pi_{sk}$ since in the long term the learner will always converge to the mastered state. The IRT form of the stationary distribution arises from maintaining a non-zero forget rate and also from characterizing some parameters as person specific and some to be item specific. 

This embedding of the two models into a unifying framework highlights their relationship. One of the reasons the BKT model has been so successful for modeling learning data is its requirement of fine-grained skills \citep{corbett1994}. The finer the skills modeled in BKT are, the closer the model comes to approximating an IRT model at equilibrium. Furthermore, this connection explains the success of extensions to BKT which allow for person-specific learning parameters \citep{pardos2011, lee2012,  yudelson2013, khajah2014A, khajah2014B}. 


By formally connecting BKT and IRT models, we obtain a new interpretation of some of the key IRT parameters. Ability is seen to be nothing other than (a function of) the probability to move from the unlearned to the learned state. That is, a person is characterized by his or her \emph{ability to learn}. Item difficulty is seen to be nothing other than (a function of) the (item specific) probability to move from the learned state back to the unlearned one, and hence modulates how long people stay in the learned state, once they've entered it.

\section{Simulation}
\begin{figure}[t]
\includegraphics[width=\textwidth,keepaspectratio]{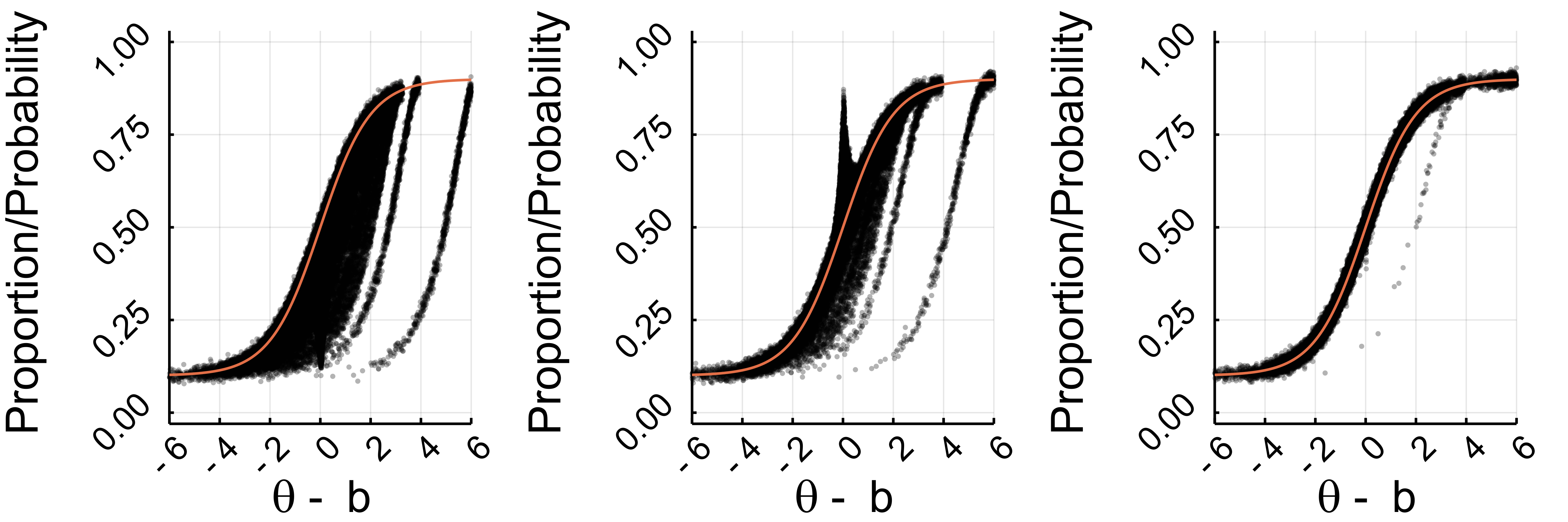}
\caption{Simulation results. The horizontal axis is the difference between ability and difficulty $\theta_p-b_i$. The vertical axis is the proportion of the 1,000 simulations in which the correct answer was submitted. The leftmost plot corresponds to number of HMM iterations equal to 2, the center plot is 5 iterations, and the rightmost plot is after 50 iterations. The 4PL IRF is superimposed in orange.}
\label{fig:sim}
\end{figure}
A small simulation illustrates the concordance between the BKT model and IRT. A total of $n=1,000$ people are simulated with learning rate $\pi_{{\ell}p}$ and a total of $m=100$ items are simulated with forgetting rate $\pi_{{\phi}i}$. The learning and forgetting parameters are drawn from a uniform distribution between 0 and 1 for $p$ in $1,\ldots,n$ and $i$ in $1,\ldots,m$. Ability and difficult are then calculated as $\theta_p=\log\pi_{{\ell}p}$ and $b_i = \log\pi_{{\phi}i}$ for $p$ in $1,\ldots,n$ and $i$ in $1,\ldots,m$.

Each person maintains a latent mastery variable for each item, $Z_{pit}$ that starts off as unmastered $Z_{pit}=0$. The states of this latent variable are then sampled according to its corresponding transition matrix determined by the learning and forgetting rates for a specific number of iterations. This number of iterations is varied and takes on values 2, 5, and 50. Each individual then submits an answer to each item. The probability a correct response is given is $1-\pi_s$ if they have mastered the item and $\pi_g$ is they have not mastered it, where $\pi_s=\pi_g=0.1$ This process is then repeated over 1,000 random replications. 

Figure \ref{fig:sim} shows the results of the simulation. The horizontal axis is the difference between ability and difficulty $\theta_p-b_i$. The vertical axis is the proportion of the 1,000 simulations in which the correct answer was submitted. The leftmost plot corresponds to number of HMM iterations equal to 2, the center plot is 5 iterations, and the rightmost plot is after 50 iterations. Each plot has superimposed on it the 4PL IRF (the orange curve). From the figure we can see that as the number of iterations increases, the distribution of the latent states approaches that of their equilibrium distribution and thus correspond to the 4PL IRF. 

\section{Future Directions}
\label{sec:future}
So where do we stand, after having explicated the formal relation between BKT and IRT models, with respect to our criticism of both for Learning Systems?
Assessment has met learning, but neither has met education thus far.
One could argue that no progress has been made. After all, integrating two models, neither one suitable for thinking about education, does not by itself lead to a model suitable for thinking about education. 

We argue that the outlook is not so bleak. The integration of BKT and IRT is a point of departure for a new research agenda for the learning sciences and psychometrics \emph{together}; an agenda aimed at factoring the role of education into the learning equation. Both the outcome of learning and the process of learning crucially depend on it. In this section we outline some key features that we believe models for learning and assessment should have. 

Let's start off with a question: Why is it that no educational system starts off in primary education by teaching long division, and then slowly working towards counting? Every systems starts off with counting, followed by addition, and slowly works its way to long division. Counting is a clear prerequisite for addition. That is, even though both learning processes may be adequately described by a BKT-IRT model, the learning processes are not independent. Leveraging such inherent dependencies is what education is all about. In \citet{doignon1985} a set theoretic approach is taken for describing these dependencies which the authors called knowledge spaces. Although the authors in this paper provide The structures and language to describe these sorts of dependency structures, they do not provide the algorithmic procedures for the assessment of knowledge. Such a modeling framework which is capable of taking these hierarchical dependencies into account are the models that come from the literature on network psychometrics.

Network psychometrics is a new conceptual model that formalizes such (mutual) dependencies. Starting from the mutualism process model of intelligence \citep{vandermaas2006}, network psychometrics has rapidly expanded and matured over the past decade. Whereas the mutualism model focussed on interactions and dependencies between unobservable quantities, the recent literature has focused on interactions between observable variables \citep{borsboom2008, borsboom2013, cramer2010}. The primary innovation in this new conception is the construction of a graphical network in which the nodes are the observable features that are mutually reinforcing based on their connections by causal relations. Recent advances in network psychometrics have highlighted the connection between several graphical models to those of psychometric models \citep{marsman2017}. Specifically, \citet{marsman2017} shows a statistical equivalence between the Ising model (a graphical model in statistical physics) and the multidimensional 2PL IRT model. 

The necessity of incorporating the dependency structure of skills can be easily seen (we wouldn't want to teach children long division before counting) however there are other less tangible and more nuanced psychological phenomena and findings that have been replicated over many studies and data sets that our models should explain. These phenomena include positive manifold (i.e. positive correlation between scores of different tests) \citep{spearman1904}, Matthew effect \citep{merton1968} (sometimes summarized by the adage that "the rich get richer and the poor get poorer"), and high predictive validity, the extent to which a score on a scale or test predicts scores on some criterion measure \citep{cronbach1955}. However, neither BKT nor IRT explain why these phenomena occur. It is not necessary in either model, for example, to have high math scores be positively correlated with high English scores. Fortunately, research in network psychometrics has identified how these well replicated phenomena are actually intrinsic to the models proposed by network psychometrics, for example  see \citet{kan2016,kovacs2016,savi2018}. 

Finally, our model should address how this network of dependencies between skills grows and changes. Learners will learn new skills or the connections between existing skills will change, typically by adding in more connections and dependencies, but this is also where forgetting can be incorporated. The paper by \citet{savi2018} introduces a network model for intelligence based on the Fortuin-Kasteleyn model. In this paper a method for studying growth dynamics is also described.  

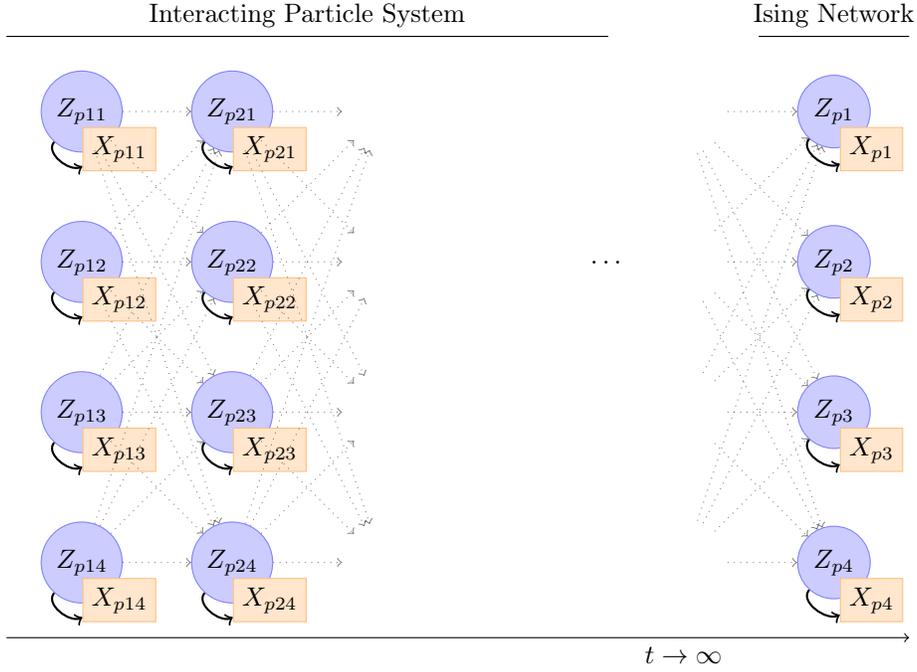
\begin{figure}[t]
\centering
\begin{tikzpicture}[]
  \draw (-1,3) -- node[auto=false, above]{Interacting Particle System} (7,3);
  \draw (9,3) -- node[auto=false, above]{Ising Network} (11,3);
  \foreach \a/\b in {1/2,2/0,3/-2,4/-4}{
    \node[state] (Z1\a) at (0,\b) {$Z_{p1\a}$};
    \node[state] (Z2\a) at (2,\b) {$Z_{p2\a}$};
    \node[state] (Z\a) at (10,\b) {$Z_{p\a}$};
    
    \node[draw=none, shape=circle] (Z3\a) at (4,\b) {$\phantom{Z_{j3\a}}$};
    \node[draw=none, shape=circle] (Zm\a) at (8,\b) {$\phantom{Z_{jm\a}}$};
    
    \node[observation] (X1\a) at (0+0.5,\b-0.5) {$X_{p1\a}$};
    \path[thick,->] (Z1\a) edge [left, in=-155, out=-132] node[above] {} (X1\a);
    
    \node[observation] (X2\a) at (2+0.5,\b-0.5) {$X_{p2\a}$};
    \path[thick,->] (Z2\a) edge [left, in=-155, out=-132] node[above] {} (X2\a);
    
    \node[observation] (X\a) at (10+0.5,\b-0.5) {$X_{p\a}$};
    \path[thick,->] (Z\a) edge [left, in=-155, out=-132] node[above] {} (X\a);
  }
  \foreach \a in {1,2,3,4}{
    \foreach \b in {1,2,3,4}{
      \draw[lightedge, opacity=0.5,->] (Z1\a) to (Z2\b);
      \draw[lightedge, opacity=0.5,->] (Z2\a) to (Z3\b);
      \draw[lightedge, opacity=0.5,->] (Zm\a) to (Z\b);
    }
  }
  \path (4,0) -- node[auto=false]{\ldots} (10,0);
  \draw [->] (-1,-5) to node[pos=0.75, below]{$t\to\infty$} (11,-5);
\end{tikzpicture}
\caption{An interacting particle system in which $Z_{pkt}$ and $X_{pkt}$ are the same as in Figure \ref{fig:hmm} but are allowed to interact with each other rather than be separate, independent models, thus producing a hidden Markov field model.}
\label{fig:hmf}
\end{figure}

We have outlined some major features that we would wish a future model to exhibit, many stemming from innovations and research done recently in network psychometrics. In light of this, it should be noted that the BKT and IRT model framework is also a subset of these network psychometrics models. It has already been noted above that IRT models such as MIRT have been shown to be related to the Ising model. It is not much of an extension from there to include BKT as well. Figure \ref{fig:hmf} illustrates this connection. On the right we have an illustration of an Ising model which is equivalent to an MIRT model. The Ising model portrayed is fully connected. The $\mathbf{a}_i$ parameters of the corresponding MIRT model are related to eigenvalue decomposition of the connectivity matrix $\Sigma$ of the Ising model, where $\sigma_{ij}$ is the interaction strength between nodes $i$ and $j$. The Ising model is difficult to evaluate numerically if there are many nodes in the network and thus it is often simulated using Monte Carlo methods such as the Metropolis algorithm or Glauber dynamics \citep{newman1999}. The structure of the Ising model along with the Monte Carlo sampling scheme used for simulating the Ising model constitutes an interacting particle system. An interacting particle system is defined by a graphical network, along with the transition probabilities for each node \citep{liggett2012}. Figure \ref{fig:hmf} is an extension of the ideas presented in Figure \ref{fig:singlehmmtoirt}. Whereas Figure \ref{fig:singlehmmtoirt} illustrates a single latent variable evolving over time to its stationary distribution, Figure \ref{fig:hmf} depicts an interconnected network of multiple variables that evolve over time to a stationary distribution that is an MIRT model (or equivalently, an Ising model).   

This generalization resolves two issues that were described with BKT and IRT as they relate to learning systems. First, there is now a placeholder for education, namely the connections between the various observable nodes and the associated strength of interaction parameters in the Ising model. Learning/education is depicted as the addition of connections between these nodes. Second, the independence of skills is not required as in BKT, but rather the dependencies between latent skills is implied by the causal relationships of the observable nodes. Whereas in IRT, the items are assumed to be conditionally independent given the latent trait, the Ising model does not require this. Indeed, the items in the Ising model exhibit dependence and this dependence is not necessarily due to a latent trait. 


We have shown how the standard models for learning and assessment are related and how they in fact fall under a larger umbrella of models in network psychometrics. The success and popularity of these models may be attributable to the powerful phenomenon that fall out of these network models. Further implications of network psychometrics on the connection between BKT and IRT need to be explored. 


\begin{thebibliography}{53}
\providecommand{\natexlab}[1]{#1}
\providecommand{\url}[1]{{#1}}
\providecommand{\urlprefix}{URL }
\expandafter\ifx\csname urlstyle\endcsname\relax
  \providecommand{\doi}[1]{DOI~\discretionary{}{}{}#1}\else
  \providecommand{\doi}{DOI~\discretionary{}{}{}\begingroup
  \urlstyle{rm}\Url}\fi
\providecommand{\eprint}[2][]{\url{#2}}

\bibitem[{Altman(2007)}]{altman2007}
Altman RM (2007) Mixed hidden markov models. Journal of the American
  Statistical Association 102(477):201--210, \doi{10.1198/016214506000001086},
  \urlprefix\url{https://doi.org/10.1198/016214506000001086},
  \eprint{https://doi.org/10.1198/016214506000001086}

\bibitem[{Beck et~al(2008)Beck, Chang, Mostow, and Corbett}]{beck2008}
Beck J, Chang Km, Mostow J, Corbett A (2008) Does help help? introducing the
  bayesian evaluation and assessment methodology. In: Intelligent Tutoring
  Systems, Springer, pp 383--394

\bibitem[{Beck and Chang(2007)}]{beck2007}
Beck JE, Chang Km (2007) Identifiability: A fundamental problem of student
  modeling. In: International Conference on User Modeling, Springer, pp
  137--146

\bibitem[{Birnbaum(1967)}]{birnbaum1967}
Birnbaum A (1967) Statistical theory for logistic mental test models with a
  prior distribution of ability. Research Bulletin No. RB-67-12, Educational
  Testing Service, Princeton, NJ

\bibitem[{Borsboom(2008)}]{borsboom2008}
Borsboom D (2008) Psychometric perspectives on diagnostic systems. Journal of
  Clinical Psychology 64(9):1089--1108, \doi{10.1002/jclp.20503},
  \urlprefix\url{http://dx.doi.org/10.1002/jclp.20503}

\bibitem[{Borsboom and Cramer(2013)}]{borsboom2013}
Borsboom D, Cramer AO (2013) Network analysis: an integrative approach to the
  structure of psychopathology. Annual review of clinical psychology 9:91--121

\bibitem[{Cen et~al(2006)Cen, Koedinger, and Junker}]{cen2006}
Cen H, Koedinger K, Junker B (2006) Learning factors analysis-a general method
  for cognitive model evaluation and improvement. In: Intelligent tutoring
  systems, Springer, vol 4053, pp 164--175

\bibitem[{Corbett and Anderson(1995)}]{corbett1994}
Corbett AT, Anderson JR (1995) Knowledge tracing: Modeling the acquisition of
  procedural knowledge. User modeling and user-adapted interaction
  4(4):253--278

\bibitem[{Cramer et~al(2010)Cramer, Waldorp, {van der Maas}, and
  Borsboom}]{cramer2010}
Cramer AOJ, Waldorp LJ, {van der Maas} HLJ, Borsboom D (2010) Comorbidity: A
  network perspective. Behavioral and Brain Sciences 33(2-3):137–150,
  \doi{10.1017/S0140525X09991567}

\bibitem[{Cronbach and Meehl(1955)}]{cronbach1955}
Cronbach LJ, Meehl PE (1955) Construct validity in psychological tests.
  Psychological bulletin 52(4):281

\bibitem[{Doignon and Falmagne(1985)}]{doignon1985}
Doignon J, Falmagne J (1985) Spaces for the assessment of knowledge.
  International Journal of Man-Machine Studies 23:175--196

\bibitem[{Doroudi and Brunskill(2017)}]{doroudi2017}
Doroudi S, Brunskill E (2017) The misidentified identifiability problem of
  bayesian knowledge tracing

\bibitem[{Ekanadham and Karklin(2017)}]{ekanadham2017}
Ekanadham C, Karklin Y (2017) T-skirt: Online estimation of student proficiency
  in an adaptive learning system. arXiv preprint arXiv:170204282

\bibitem[{Embretson(1991)}]{embretson1991}
Embretson SE (1991) A multidimensional latent trait model for measuring
  learning and change. Psychometrika 56(3):495--515

\bibitem[{Gowda et~al(2010)Gowda, Rowe, Baker, Chi, and Koedinger}]{gowda2010}
Gowda SM, Rowe JP, Baker RS, Chi M, Koedinger KR (2010) Improving models of
  slipping, guessing, and moment-by-moment learning with estimates of skill
  difficulty. In: Educational Data Mining 2011

\bibitem[{Green(1950)}]{green50A}
Green B (1950) A general solution for the latent class model of latent
  structure analysis. Research Bulletin No. RB-50-38, Educational Testing
  Service, Princeton, NJ

\bibitem[{Gweon et~al(2015)Gweon, Lee, Dorsey, Tinker, Finzer, and
  Damelin}]{gweon2015}
Gweon GH, Lee HS, Dorsey C, Tinker R, Finzer W, Damelin D (2015) Tracking
  student progress in a game-like learning environment with a monte carlo
  bayesian knowledge tracing model. In: Proceedings of the Fifth International
  Conference on Learning Analytics And Knowledge, ACM, pp 166--170

\bibitem[{Kan et~al(2016)Kan, {van der Maas}, and Kievit}]{kan2016}
Kan KJ, {van der Maas} HLJ, Kievit RA (2016) Process overlap theory: Strengths,
  limitations, and challenges. Psychological Inquiry 27(3):220--228,
  \doi{10.1080/1047840X.2016.1182000},
  \urlprefix\url{https://doi.org/10.1080/1047840X.2016.1182000},
  \eprint{https://doi.org/10.1080/1047840X.2016.1182000}

\bibitem[{Khajah et~al(2014{\natexlab{a}})Khajah, Wing, Lindsey, and
  Mozer}]{khajah2014B}
Khajah M, Wing R, Lindsey R, Mozer M (2014{\natexlab{a}}) Integrating
  latent-factor and knowledge-tracing models to predict individual differences
  in learning. In: Educational Data Mining 2014

\bibitem[{Khajah et~al(2016)Khajah, Lindsey, and Mozer}]{khajah2016}
Khajah M, Lindsey RV, Mozer MC (2016) How deep is knowledge tracing? arXiv
  preprint arXiv:160402416

\bibitem[{Khajah et~al(2014{\natexlab{b}})Khajah, Huang, Gonz\'{a}lez-Brenes,
  Mozer, and Brusilovsky}]{khajah2014A}
Khajah MM, Huang Y, Gonz\'{a}lez-Brenes JP, Mozer MC, Brusilovsky P
  (2014{\natexlab{b}}) Integrating knowledge tracing and item response theory:
  A tale of two frameworks. In: Proceedings of Workshop on Personalization
  Approaches in Learning Environments (PALE 2014) at the 22th International
  Conference on User Modeling, Adaptation, and Personalization, University of
  Pittsburgh, pp 7--12

\bibitem[{Koedinger et~al(2013)Koedinger, Stamper, McLaughlin, and
  Nixon}]{koedinger2013}
Koedinger KR, Stamper JC, McLaughlin EA, Nixon T (2013) Using data-driven
  discovery of better student models to improve student learning. In: Lane HC,
  Yacef K, Mostow J, Pavlik P (eds) Artificial Intelligence in Education,
  Springer Berlin Heidelberg, Berlin, Heidelberg, pp 421--430

\bibitem[{Kovacs and Conway(2016)}]{kovacs2016}
Kovacs K, Conway ARA (2016) Process overlap theory: A unified account of the
  general factor of intelligence. Psychological Inquiry 27(3):151--177,
  \doi{10.1080/1047840X.2016.1153946},
  \urlprefix\url{https://doi.org/10.1080/1047840X.2016.1153946},
  \eprint{https://doi.org/10.1080/1047840X.2016.1153946}

\bibitem[{Lee and Brunskill(2012)}]{lee2012}
Lee JI, Brunskill E (2012) The impact on individualizing student models on
  necessary practice opportunities. International Educational Data Mining
  Society

\bibitem[{Liggett(2012)}]{liggett2012}
Liggett TM (2012) Interacting particle systems, vol 276. Springer Science \&
  Business Media

\bibitem[{Lord(1977)}]{lord1977}
Lord F (1977) Practical applications of item characteristic curve theory.
  Journal of Educational Measurement 14(2):117--138,
  \doi{10.1111/j.1745-3984.1977.tb00032.x},
  \urlprefix\url{http://dx.doi.org/10.1111/j.1745-3984.1977.tb00032.x}

\bibitem[{Lord(1951)}]{lord1951}
Lord FM (1951) A theory of test scores and their relation to the trait
  measured. Research Bulletin No. RB-51-13, Educational Testing Service,
  Princeton, NJ

\bibitem[{Maclellan et~al(2015)Maclellan, Liu, and Koedinger}]{maclellan2015}
Maclellan C, Liu R, Koedinger K (2015) Accounting for slipping and other false
  negatives in logistic models of student learning. In: Proceedings for the 8th
  International Conference on Educational Data Mining

\bibitem[{Marsman et~al(2018)Marsman, Borsboom, Kruis, Epskamp, {van Bork},
  Waldorp, {van der Maas}, and Maris}]{marsman2017}
Marsman M, Borsboom D, Kruis J, Epskamp S, {van Bork} R, Waldorp LJ, {van der
  Maas} HLJ, Maris G (2018) An introduction to network psychometrics: Relating
  ising network models to item response theory models. Multivariate Behavioral
  Research 53(1):15--35, \doi{10.1080/00273171.2017.1379379},
  \urlprefix\url{https://doi.org/10.1080/00273171.2017.1379379}, pMID:
  29111774, \eprint{https://doi.org/10.1080/00273171.2017.1379379}

\bibitem[{Martin and Quinn(2002)}]{martin2002}
Martin AD, Quinn KM (2002) Dynamic ideal point estimation via markov chain
  monte carlo for the us supreme court, 1953--1999. Political Analysis
  10(2):134--153

\bibitem[{Merton(1968)}]{merton1968}
Merton RK (1968) The matthew effect in science: The reward and communication
  systems of science are considered. Science 159(3810):56--63

\bibitem[{Nedungadi and Remya(2015)}]{nedungadi2015}
Nedungadi P, Remya M (2015) Incorporating forgetting in the personalized,
  clustered, bayesian knowledge tracing (pc-bkt) model. In: Cognitive Computing
  and Information Processing (CCIP), 2015 International Conference on, IEEE, pp
  1--5

\bibitem[{Newman and Barkema(1999)}]{newman1999}
Newman M, Barkema G (1999) Monte Carlo Methods in Statistical Physics chapter
  1-4. Oxford University Press: New York, USA

\bibitem[{Pardos and Heffernan(2011)}]{pardos2011}
Pardos Z, Heffernan N (2011) Kt-idem: introducing item difficulty to the
  knowledge tracing model. User Modeling, Adaption and Personalization pp
  243--254

\bibitem[{Pavlik et~al(2009)Pavlik, Cen, and Koedinger}]{pavlik2009}
Pavlik PI, Cen H, Koedinger KR (2009) Performance factors analysis --a new
  alternative to knowledge tracing. In: Proceedings of the 2009 Conference on
  Artificial Intelligence in Education: Building Learning Systems That Care:
  From Knowledge Representation to Affective Modelling, IOS Press, Amsterdam,
  The Netherlands, The Netherlands, pp 531--538,
  \urlprefix\url{http://dl.acm.org/citation.cfm?id=1659450.1659529}

\bibitem[{Qiu et~al(2010)Qiu, Qi, Lu, Pardos, and Heffernan}]{qiu2010}
Qiu Y, Qi Y, Lu H, Pardos Z, Heffernan N (2010) Does time matter? modeling the
  effect of time with bayesian knowledge tracing. In: Educational Data Mining
  2011

\bibitem[{Rasch(1960)}]{rasch1960}
Rasch G (1960) Probabilistic Models for Some Intelligence and Attainment Tests.
  Studies in mathematical psychology, Danmarks Paedagogiske Institut,
  \urlprefix\url{https://books.google.com/books?id=aB9qLgEACAAJ}

\bibitem[{Reckase(1972)}]{reckase1972}
Reckase MD (1972) Development and application of a multivariate logistic latent
  trait model. PhD thesis, Syracuse University

\bibitem[{Ross(2014)}]{ross2014}
Ross S (2014) Introduction to Probability Models. Elsevier Science,
  \urlprefix\url{https://books.google.com/books?id=cehTngEACAAJ}

\bibitem[{{Sao Pedro} et~al(2013){Sao Pedro}, Baker, and Gobert}]{sao2013}
{Sao Pedro} M, Baker R, Gobert J (2013) Incorporating scaffolding and tutor
  context into bayesian knowledge tracing to predict inquiry skill acquisition.
  In: Educational Data Mining 2013

\bibitem[{Savi et~al(2018)Savi, Marsman, {van der Maas}, and Maris}]{savi2018}
Savi AO, Marsman M, {van der Maas} HLJ, Maris G (2018) The wiring of
  intelligence. PsyArXiv preprint \doi{10.31234/osf.io/32wr8}

\bibitem[{Schultz and Tabor(2013)}]{schultz2013}
Schultz S, Tabor T (2013) Revisiting and extending the item difficulty effect
  model. In: In Proceedings of the 1st Workshop on Massive Open Online Courses
  at the 16th Annual Conference on Artificial Intelligence in Education, pp
  33--40

\bibitem[{Spearman(1904)}]{spearman1904}
Spearman C (1904) "general intelligence," objectively determined and measured.
  The American Journal of Psychology 15(2):201--292,
  \urlprefix\url{http://www.jstor.org/stable/1412107}

\bibitem[{Studer(2012)}]{studer2012}
Studer C (2012) Incorporating learning into the cognitive assessment framework.
  PhD thesis, Carnegie Mellon University

\bibitem[{Sympson(1978)}]{sympson1978}
Sympson JB (1978) A model for testing with multidimensional items. In:
  Proceedings of the 1977 computerized adaptive testing conference, University
  of Minneapolis, Department of Psychology, Psychometric Methods Program
  Minneapolis, MN, 00014

\bibitem[{{van de Sande}(2013)}]{vandesande2013}
{van de Sande} B (2013) Properties of the bayesian knowledge tracing model.
  Journal of Educational Data Mining 5(2):1

\bibitem[{{van der Linden}(2016--2018)}]{irthandbook}
{van der Linden} W (2016--2018) Handbook of item response theory

\bibitem[{{van der Maas} et~al(2006){van der Maas}, Dolan, Grasman, Wicherts,
  Huizenga, and Raijmakers}]{vandermaas2006}
{van der Maas} HL, Dolan CV, Grasman RP, Wicherts JM, Huizenga HM, Raijmakers
  ME (2006) A dynamical model of general intelligence: the positive manifold of
  intelligence by mutualism. Psychological review 113(4):842

\bibitem[{{van Rijn}(2008)}]{rijn2008}
{van Rijn} PW (2008) Categorical time series in psychological measurement. PhD
  thesis, Universiteit van Amsterdam

\bibitem[{Wang et~al(2013)Wang, Berger, Burdick et~al}]{WangXJ2013}
Wang X, Berger JO, Burdick DS, et~al (2013) Bayesian analysis of dynamic item
  response models in educational testing. The Annals of Applied Statistics
  7(1):126--153

\bibitem[{Weng and Coad(2017)}]{weng2017}
Weng RCH, Coad DS (2017) Real-time bayesian parameter estimation for item
  response models. Bayesian Analysis

\bibitem[{Xu and Mostow(2010)}]{xu2010}
Xu Y, Mostow J (2010) Using logistic regression to trace multiple sub-skills in
  a dynamic bayes net. In: Educational Data Mining 2011

\bibitem[{Yudelson et~al(2013)Yudelson, Koedinger, and Gordon}]{yudelson2013}
Yudelson MV, Koedinger KR, Gordon GJ (2013) Individualized bayesian knowledge
  tracing models. In: International Conference on Artificial Intelligence in
  Education, Springer, pp 171--180

\end{thebibliography}
\end{document}